\documentclass[prd,superscriptaddress,amsfonts,amssymb,amsmath,showpacs,twocolumn]{revtex4-2}
\usepackage{bm}
\usepackage{amsfonts}
\usepackage{latexsym}
\usepackage[latin1]{inputenc}
\usepackage{graphicx}
\usepackage{amsmath}
\usepackage{palatino}
\usepackage{ragged2e}
\usepackage{mathpazo}
\usepackage{textcomp}
\usepackage{float}
\usepackage{booktabs}
\usepackage{dcolumn}
\usepackage{multirow}
\usepackage{hyperref}
\hypersetup{colorlinks,citecolor=blue}
\usepackage{amsmath}
\usepackage{xcolor}
\usepackage{orcidlink}
\usepackage[caption=false]{subfig}
\usepackage{commath}
\def\jnl@style{\it}
\def\aaref@jnl#1{{\jnl@style#1}}

\def\aaref@jnl#1{{\jnl@style#1}}

\def\aj{\aaref@jnl{AJ}}                   
\def\apj{\aaref@jnl{ApJ}}                 
\def\apjl{\aaref@jnl{ApJ}}                
\def\apjs{\aaref@jnl{ApJS}}               
\def\apss{\aaref@jnl{Ap\&SS}}             
\def\aap{\aaref@jnl{A\&A}}                
\def\aapr{\aaref@jnl{A\&A~Rev.}}          
\def\aaps{\aaref@jnl{A\&AS}}              
\def\mnras{\aaref@jnl{Mon.~Not.~Roy.~Astron.~Soc.}}             
\def\prd{\aaref@jnl{Phys.~Rev.~D}}        
\def\prc{\aaref@jnl{Phys.~Rev.~C}}  
\def\prl{\aaref@jnl{Phys.~Rev.~Lett.}}    
\def\qjras{\aaref@jnl{QJRAS}}             
\def\skytel{\aaref@jnl{S\&T}}             
\def\ssr{\aaref@jnl{Space~Sci.~Rev.}}     
\def\zap{\aaref@jnl{ZAp}}                 
\def\nat{\aaref@jnl{Nature}}              
\def\aplett{\aaref@jnl{Astrophys.~Lett.}} 
\def\apspr{\aaref@jnl{Astrophys.~Space~Phys.~Res.}} 
\def\physrep{\aaref@jnl{Phys.~Rep.}}      
\def\physscr{\aaref@jnl{Phys.~Scr}}       
\def\commat{\aaref@jnl{Comm.~Math.~Phys.}}              
\def\science{\aaref@jnl{Science}}               
\def\cqg{\aaref@jnl{Classical Quant.~Grav.}}            
\def\jpcs{\aaref@jnl{JPCS}}                                     
\def\ijmpd{\aaref@jnl{Int.~J.~Mod.~Phys.~D}}                    
\def\grg{\aaref@jnl{Gen.~Relat.~Gravit.}}               
\def\rpp{\aaref@jnl{Rep.~Prog.~Phys.}}          
\def\npa{\aaref@jnl{Nucl.~Phys.~A}}        
\def\lrr{\aaref@jnl{Living Rev.~Rel.}}                   
\def\jcap{\aaref@jnl{J.~Cosmology Astropart.~Phys.}}    
\def\rmp{\aaref@jnl{Rev.~Mod.~Phys.}}   
\def\epjc{\aaref@jnl{Eur.~Phys.~J.~C}}

\allowdisplaybreaks[1]

\begin{document}

\color{black}       

\title{Anisotropic $f\left( Q\right) $ gravity model with bulk viscosity}

\author{M. Koussour\orcidlink{0000-0002-4188-0572}}
\email[Email: ]{pr.mouhssine@gmail.com}
\affiliation{Quantum Physics and Magnetism Team, LPMC, Faculty of Science Ben
M'sik,\\
Casablanca Hassan II University,
Morocco.} 

\author{S. H. Shekh\orcidlink{0000-0003-4545-1975}}
\email[Email: ]{da\_salim@rediff.com}
\affiliation{Department of Mathematics. S. P. M. Science and Gilani Arts Commerce College, Ghatanji, Dist. Yavatmal, Maharashtra-445301, India.}

\author{M. Bennai\orcidlink{0000-0002-7364-5171}}
\email[Email: ]{mdbennai@yahoo.fr}
\affiliation{Quantum Physics and Magnetism Team, LPMC, Faculty of Science Ben
M'sik,\\
Casablanca Hassan II University,
Morocco.} 
\affiliation{Lab of High Energy Physics, Modeling and Simulations, Faculty of
Science,\\
University Mohammed V-Agdal, Rabat, Morocco.}

\author{N. Myrzakulov\orcidlink{0000-0001-8691-9939}}
\email[Email: ]{nmyrzakulov@gmail.com}
\affiliation{L. N. Gumilyov Eurasian National University, Astana 010008,
Kazakhstan.}

\date{\today}
\begin{abstract}

This study investigates the dynamics of a spatially homogeneous and anisotropic LRS Bianchi type-I universe with viscous fluid in the framework of $f(Q)$ symmetric teleparallel gravity. We assume a linear form for $f(Q)$ and introduce hypotheses regarding the relationship between the expansion and shear scalars, as well as the Hubble parameter and bulk viscous coefficient. The model is constrained using three observational datasets: the Hubble dataset (31 data points), the Pantheon SN dataset (1048 data points), and the BAO dataset (6 data points). The calculated cosmological parameters indicate expected behavior for matter-energy density and bulk viscous pressure, supporting the universe's accelerating expansion. Diagnostic tests suggest that the model aligns with a $\Lambda$CDM model in the far future and resides in the quintessence region. These findings are consistent with recent observational data and contribute to our understanding of cosmic evolution within the context of modified gravity and bulk viscosity.


\end{abstract}

\maketitle

\section{Introduction}\label{sec1}

The visible Universe, including the Earth, Sun, stars, and galaxies, is primarily composed of protons, neutrons, and electrons, collectively known as ordinary matter or baryonic matter. This baryonic matter accounts for less than 5\% of the total density of the Universe. In contrast, the remaining 95\% of the Universe, as indicated by studies involving high-redshift Supernovae (SNe) \cite{ref1, ref2, ref3}, Wilkinson Microwave Anisotropy Probe (WMAP) data \cite{ref4, ref5}, Cosmic Microwave Background (CMB) peaks \cite{ref6}, and Baryon Acoustic Oscillations (BAOs) \cite{ref7}, is comprised of other forms of energy and matter. These forms are currently unidentified and are referred to as Dark Energy (DE) and Dark Matter (DM). DE is an enigmatic component responsible for the observed accelerating expansion of the Universe. It possesses a positive energy density and negative pressure, manifesting as a large-scale repulsive force capable of counteracting the gravitational force that binds the various constituents of the Universe. One explanation for DE is that it corresponds to the cosmological constant $\Lambda$ that Einstein introduced into his General Relativity (GR) equations to achieve a stable Universe. However, this idea gave rise to other issues, such as the cosmic coincidence problem and the fine-tuning problem \cite{S.W.,E.J.}. Consequently, more appealing dynamical models have emerged based on the concept of modifying the matter content of the Universe. These models include quintessence, k-essence, Chapylygin gas, holographic DE, running vacuum models, and others \cite{ref8, ref9, ref10, ref11,V1,V2,V3,V4,V5}.

Recently, with a growing interest among researchers in addressing the issue of cosmic acceleration, various alternative approaches have emerged under the umbrella of modified theories of gravity (MTG). These theories seek to amend the standard Einstein-Hilbert action by replacing the Ricci scalar curvature, denoted as $R$, with arbitrary functions of this scalar, such as $f(R)$ \cite{ref12, ref13}. In addition, researchers have explored alternative theories involving other physical quantities, such as $f(R,\mathcal{T})$ gravity (where $R$ is the Ricci scalar and $\mathcal{T}$ represents the trace of the energy-momentum tensor) \cite{ref14, ref15, ref16}, $f(G)$ gravity (where $G$ signifies the Gauss-Bonnet invariant) \cite{ref17, ref18, ref19}, $f(T)$ gravity (where $T$ corresponds to the torsion scalar) \cite{ref20, ref21, ref22}, and $f(Q)$ gravity (where $Q$ is the non-metricity scalar), among others. In this study, we will investigate a cosmological model aimed at explaining cosmic acceleration within the framework of $f(Q)$ symmetric teleparallel gravity (STG), as originally proposed by Jim\'enez et al. \cite{ref23}. In this theory, the non-metricity scalar $Q$ plays a crucial role in governing gravitational interactions. Various aspects related to this theory have been explored by researchers, including energy conditions \cite{ref24}, cosmography \cite{ref25}, spherically symmetric configurations \cite{ref26}, the signature of $f(Q)$ gravity \cite{ref27}, the growth index of matter perturbations \cite{ref28}, coupling with matter \cite{ref29}, quantum cosmology for a polynomial $f(Q)$ model \cite{ref30}, the geodesic deviation equation \cite{ref31}, and the isotropization of LRS Bianchi-I Universes \cite{ref32}. Furthermore, several researchers have contributed to discussions on these topics within the framework of $f(Q)$ gravity \cite{ref33, ref34, ref35, ref36, ref37, fQ1, fQ2, fQ3, fQ4}.

In cosmology, many researchers have used perfect fluids in models to tackle scientific puzzles like cosmic acceleration, DE, DM, and primordial singularities. Recent observations hint that cosmic acceleration might be due to exotic energy with negative pressure. Building on this, our paper aims to develop a cosmological model without invoking DE. Instead, we use a more realistic approach by incorporating a viscous fluid. Previously, in the study of the inflationary epoch in the early universe, bulk viscosity has been proposed in the literature as a mechanism that does not require DE \cite{ref38,ref39, ref40,ref41}. Therefore, it is reasonable to consider that bulk viscosity could be responsible for the current accelerated expansion of the universe. In recent times, several authors have attempted to explain late-time acceleration using bulk viscosity without the need for DE or a cosmological constant \cite{ref42,ref43,ref44}. Theoretically, deviations from local thermodynamic stability can give rise to bulk viscosity, but a detailed mechanism for its formation remains elusive \cite{ref38}. In cosmology, when the matter content of the universe expands or contracts too rapidly as a cosmological fluid, an effective pressure is generated to restore the system to thermal stability. Bulk viscosity is the manifestation of this effective pressure \cite{Wilson,Okumura}. Recently, Ren et al. \cite{ref45} proposed a viscosity coefficient that depends on velocity and acceleration to achieve an accelerating expanding Universe, given by
\begin{equation}
\zeta =\zeta _{0}+\zeta _{1}\left( \frac{\overset{.}{a}}{a}\right) +\zeta
_{2}\left( \frac{\overset{..}{a}}{\overset{.}{a}}\right) =\zeta _{0}+\zeta
_{1}H+\zeta _{2}\left( \frac{\overset{.}{H}}{H}+H\right) .  \label{eqn1}
\end{equation}%
where $\zeta _{0}$, $\zeta _{1}$, and $\zeta _{2}$ are constants. As seen from the equation above, the bulk viscosity coefficient comprises a linear combination of three terms: the first is a constant, the second is proportional to the Hubble parameter $\left( H=\frac{\overset{.}{a}}{a}\right)$, indicating its dependence on velocity, and the third is proportional to $\left( \frac{\overset{..}{a}}{\overset{.}{a}}\right)$, signifying its dependence on acceleration.

Motivated by the preceding discussion and recent Planck results \cite{ref46}, which revealed defects in the CMB attributed to quantum fluctuations during the inflationary era, this study delves into the examination of the spatially homogeneous and anisotropic LRS Bianchi type-I (B-I) Universe within the framework of STG. To address the field equations within the context of STG, we employ the following two hypotheses: (i) establishing a relationship between the directional Hubble parameters $H_{x}=nH_{y}$ derived from the proportionality between the expansion scalar $\theta$ and the shear scalar $\sigma$, i.e., $\theta^{2} \propto \sigma^{2}$, and (ii) establishing a relationship between the average Hubble parameter $H$ and the bulk viscous coefficient as described in Eq. (\ref{eqn1}). The study explores various cosmological parameters, including the deceleration parameter, equation of state (EoS) parameter, statefinder, and $Om(z)$ diagnostic parameters, in the context of this model. The structure of this paper is as follows: In Sec. \ref{sec2}, we present the fundamental equations of $f(Q)$ gravity. Sec. \ref{sec3} introduces the B-I Universe influenced by bulk viscous fluid matter, presenting the exact solution for the Hubble parameter. Subsequently, in Sec. \ref{sec4}, we perform an analysis of observational data to determine the best-fit values for the parameters, utilizing the Hubble dataset with 31 points, the Pantheon dataset comprising 1048 samples, and the BAO sample. Sec. \ref{sec5} is dedicated to the discussion of various cosmological parameters, including the deceleration parameter, EoS parameter, statefinder, and $Om(z)$ diagnostic parameters, analyzing their behavior concerning the redshift $z$. The final section provides a summary of the results and the conclusions drawn from this study.
 
\section{Basic equations of $f\left( Q\right) $ gravity}
\label{sec2}

The action for $f\left( Q\right) $ theory of gravity is given by \cite%
{ref23}
\begin{equation}
S=\int \left[ \frac{1}{2 \kappa}f(Q)+L_{m}\right] d^{4}x\sqrt{-g},  \label{eqn2}
\end{equation}%
where $f(Q)$ represents an arbitrary function of the non-metricity scalar $Q$. Here, $g$ denotes the determinant of the metric tensor $g_{\mu \nu }$, i.e., $g=\det \left( g_{\mu \nu }\right)$, and $L_{m}$ represents the conventional matter Lagrangian. The non-metricity scalar $Q$ is determined as
\begin{equation}
Q=-g^{\mu \nu }\left( {L^{\alpha }}_{\beta \mu }{L^{\beta }}_{\nu \alpha }-{%
L^{\alpha }}_{\beta \alpha }{L^{\beta }}_{\mu \nu }\right),
\label{eqn3}
\end{equation}%
where the deformation tensor ${L^{\gamma }}_{\mu \nu }$ is defined as,
\begin{eqnarray}
{L^{\gamma }}_{\mu \nu }=\frac{1}{2}g^{\gamma \sigma }\left( -Q_{\mu \sigma \nu }-Q_{\nu \sigma \mu
}+Q_{\sigma \mu \nu }\right)={L^{\gamma }}_{\nu \mu }.
\end{eqnarray}

The non-metricity tensor is defined in the form:
\begin{equation}
Q_{\gamma \mu \nu }=\nabla _{\gamma }g_{\mu \nu }.  \label{eqn5}
\end{equation}%

The trace of the non-metricity tensor is obtained as follows:
\begin{equation}
Q_{\gamma }=Q_{\gamma }{}^{\mu }{}_{\mu }\text{ \ \ and \ \ }\tilde{Q}%
_{\gamma }=Q^{\mu }{}_{\gamma \mu }.   \label{eqn6}
\end{equation}

In addition, we define the superpotential tensor:
\begin{equation}
4P^{\gamma }{}_{\mu \nu }=-Q^{\gamma }{}_{\mu \nu }+2Q_{(\mu }{}^{\gamma
}{}_{\nu )}+(Q^{\gamma }-\tilde{Q}^{\gamma })g_{\mu \nu }-\delta _{(\mu
}^{\gamma }Q_{\nu )}.
\label{eqn7}
\end{equation}%

Using this definition, the non-metricity scalar is expressed as
\begin{equation}
Q=-Q_{\gamma \mu \nu }P^{\gamma \mu \nu }.  \label{eqn8}
\end{equation}

Now, the energy-momentum tensor for matter is defined by the following mathematical relation:
\begin{equation}
\mathcal{T}_{\mu\nu}=\frac{-2}{\sqrt{-g}}\frac{\delta \left( \sqrt{-g}L_{m}\right) }{%
\delta g^{\mu \nu }}.  \label{eqn9}
\end{equation}

The field equations for $f(Q)$ gravity are obtained by varying the action $(S)$ in Eq. (\ref{eqn2}) with respect to the metric tensor $g_{\mu \nu }$,
\begin{equation}
\begin{split}
&\frac{2}{\sqrt{-g}}\nabla_\gamma (\sqrt{-g}f_Q P^\gamma\:_{\mu\nu})+ \frac{1}{2}g_{\mu\nu}f\\
&+f_Q(P_{\mu\gamma\beta}Q_\nu\:^{\gamma\beta} - 2Q_{\gamma\beta\mu}P^{\gamma\beta}\:_\nu) = -\kappa \mathcal{T}_{\mu\nu},  \label{eqn10}
\end{split}
\end{equation}
where $f_{Q}=\frac{df}{dQ}$, $\nabla _{\gamma }$ represents the covariant derivative, and for simplicity, we adopt natural units $\left( \kappa=8\pi G=1\right) $. In addition, we can also perform a variation of (\ref{eqn2}) with respect to the connection, leading to the following result:
\begin{equation}
\nabla_\mu \nabla_\nu (\sqrt{-g}f_Q P^{\mu\nu}\:_\gamma) =  0.  \label{eqn11}
\end{equation}

\section{LRS Bianchi type-I Universe with bulk viscosity}
\label{sec3}

In the present discussion, our focus is on the spatially homogeneous and anisotropic LRS B-I Universe. This cosmological model is a direct generalization of the flat FLRW Universe and is described by the following metric form:
\begin{equation}
ds^{2}=-dt^{2}+A^{2}(t)dx^{2}+B^{2}(t)\left( dy^{2}+dz^{2}\right) ,
\label{eqn12}
\end{equation}%
Here, $t$ represents cosmic time, and the scale factors $A(t)$ and $B(t)$ characterize the expansion or contraction of the Universe in different spatial directions. A flat FLRW space-time can be achieved by setting $A(t) = B(t) = a(t)$. This anisotropic LRS B-I Universe provides a valuable framework for exploring various cosmological phenomena, offering insights into the behavior of the cosmos beyond the simplifications of the homogeneous and isotropic FLRW models. In the following sections, we will delve into the dynamics and implications of this intriguing cosmological scenario within the context of $f(Q)$ gravity. The corresponding non-metricity scalar is given by:
\begin{equation}
Q=-2\left( \frac{\overset{.}{B}}{B}\right) ^{2}-4\frac{\overset{.}{A}}{A}%
\frac{\overset{.}{B}}{B}.  \label{eqn13}
\end{equation}

The inclusion of viscous effects in the cosmic fluid content can be interpreted as an effort to enhance the precision of its description, introducing a departure from its idealized properties. This viscous contribution negatively influences the total pressure, playing a role in propelling the cosmic late-time acceleration \cite{ref43,ref44,ref45}. The energy-momentum tensor describing a Universe filled with viscous content can be expressed as
\begin{equation}
\mathcal{T}_{\mu\nu}=\left( \rho +p_{v}\right) u_{\mu }u_{\nu }+p_{v}h_{\mu \nu },
\label{eqn14}
\end{equation}%

In this context, we introduce the metric tensor $h_{\mu \nu}=g_{\mu \nu }+u_{\mu }u_{\nu}$, where $\rho$ is the usual matter energy density and $p_{v}$ is the pressure of the bulk viscous fluid and is defined as $p_{v}=p-3\zeta H$. Here, $p$ signifies the pressure of the perfect fluid, and the coefficient of bulk viscosity, $\zeta$, is typically a function of the Hubble parameter $H$ and its derivatives, as indicated in Eq. (\ref{eqn1}). The four-velocity vector $u^{\mu }$ is assumed to satisfy $u^{\mu }u_{\mu }=-1$. Currently, the Universe is primarily composed of non-relativistic matter (dust), which leads to $p_{v}=-3\zeta H$. With the bulk viscous fluid as the dominant matter component, the corresponding field equations for the B-I Universe can be derived as follows \cite{ref32}:
\begin{widetext}
\begin{equation}
\frac{f}{2}+f_{Q}\left[ 4\frac{\overset{.}{A}}{A}\frac{\overset{.}{B}}{B}%
+2\left( \frac{\overset{.}{B}}{B}\right) ^{2}\right] =\rho ,  \label{eqn15}
\end{equation}
\begin{equation}
\frac{f}{2}-f_{Q}\left[ -2\frac{\overset{.}{A}}{A}\frac{\overset{.}{B}}{B}-2%
\frac{\overset{..}{B}}{B}-2\left( \frac{\overset{.}{B}}{B}\right) ^{2}\right]
+2\frac{\overset{.}{B}}{B}\overset{.}{Q}f_{QQ}=-p_{v},  \label{eqn16}
\end{equation}
\begin{equation}
\frac{f}{2}-f_{Q}\left[ -3\frac{\overset{.}{A}}{A}\frac{\overset{.}{B}}{B}-%
\frac{\overset{..}{A}}{A}-\frac{\overset{..}{B}}{B}-\left( \frac{\overset{.}{%
B}}{B}\right) ^{2}\right] +\left( \frac{\overset{.}{A}}{A}+\frac{\overset{.}{%
B}}{B}\right) \overset{.}{Q}f_{QQ}=-\left( p_{v}+\delta \rho \right).
\label{eqn17}
\end{equation}%
\end{widetext}

Here, $\delta$ is referred to as the skewness parameter, quantifying deviations from the EoS parameter along the $y$ and $z$ directions. In addition, the notation $(\dot{})$ represents a derivative with respect to cosmic time $t$. The field equations presented in Eq. (\ref{eqn8}) through Eq. (\ref{eqn10}) can be expressed in terms of the mean Hubble parameter and directional Hubble parameters as follows:
\begin{equation}
\frac{f}{2}-Qf_{Q}=\rho ,  \label{eqn18}
\end{equation}
\begin{equation}
\frac{f}{2}+2\frac{\partial }{\partial t}\left[ H_{y}f_{Q}\right]
+6Hf_{Q}H_{y}=-p_{v},  \label{eqn19}
\end{equation}
\begin{equation}
\frac{f}{2}+\frac{\partial }{\partial t}\left[ f_{Q}\left(
H_{x}+H_{y}\right) \right] +3Hf_{Q}\left( H_{x}+H_{y}\right) =-\left(
p_{v}+\delta \rho \right).  \label{eqn20}
\end{equation}%

In the derivation,  we utilized $\frac{\partial }{\partial t}\left( \frac{\overset{.}{A}}{A}%
\right) =\frac{\overset{..}{A}}{A}-\left( \frac{\overset{.}{A}}{A}\right)
^{2}$ and $Q=-2H_{y}^{2}-4H_{x}H_{y}$. Here, $H=\frac{\overset{.}{a}}{a}=%
\frac{1}{3}\left( H_{x}+2H_{y}\right) $ represents the average Hubble parameter, while $H_{x}=\frac{\overset{.}{A}}{A}$, $H_{y}=H_{z}=\frac{\overset{.}{B}}{B}$ denote the directional Hubble parameters along the $x$, $y$, and $z$ axes, respectively.

In this paper, inspired by the work presented in \cite{ref47}, we consider the following linear $f(Q)$ model, which is characterized by a functional form of the non-metricity scalar $Q$ given by
\begin{equation}
f\left( Q\right) =\alpha Q,\text{ \ \ }\alpha \neq 0,  \label{eqn21}
\end{equation}
where $\alpha$ is a constant parameter. The choice of a linear model for $f(Q)$ can be motivated by several considerations within the context of modified gravity theories \cite{Solanki1,Solanki2, Solanki3}. Also, the linear model of $f(Q)$ gravity is equivalent to GR with a different gravitational constant.

Upon incorporating the final constraint, the field equations (\ref{eqn18})-(%
\ref{eqn19}) form a system of three differential equations involving five unknowns. Consequently, to obtain exact solutions for the field equations, an additional constraint is required. In this context, we impose the physical condition that the expansion scalar $\theta $ is directly proportional to the shear scalar $\sigma$, i.e., $\theta^{2} \propto \sigma ^{2}$. This condition leads to the relation:
\begin{equation}
H_{x}=nH_{y},  \label{eqn22}
\end{equation}%
where $n\left( \neq 0,1\right) $ represents an arbitrary real number. The physical rationale behind this assumption is grounded in observations of the velocity-redshift relation for extragalactic sources, suggesting that the Hubble expansion of the Universe may tend toward isotropy when the ratio $\frac{\sigma }{\theta }$ remains constant \cite{ref48}. This condition has been applied in various studies \cite{ref32, ref36, ref37}. Utilizing Eq. (\ref{eqn22}), we can derive expressions for the average Hubble parameter and non-metricity scalar in terms of $H_{y}$ as follows:
\begin{equation}
H=\frac{\left( n+2\right) }{3}H_{y},  \label{eqn23}
\end{equation}
\begin{equation}
Q=-2\left( 1+2n\right) H_{y}^{2}.  \label{eqn24}
\end{equation}

Thus, by employing Eqs. (\ref{eqn21}), (\ref{eqn23}), and (\ref{eqn24}), the field equations take the following form:
\begin{equation}
\alpha \left( 1+2n\right) H_{y}^{2}=\rho ,  \label{eqn25}
\end{equation}
\begin{equation}
3\alpha H_{y}^{2}+2\alpha \overset{.}{H}_{y}=-p_{v},  \label{eqn26}
\end{equation}
\begin{equation}
\alpha \left( n^{2}+n+1\right) H_{y}^{2}+\alpha \left( n+1\right) \overset{.}%
{H}_{y}=-\left( p_{v}+\delta \rho \right) .  \label{eqn27}
\end{equation}

By combining Eqs. (\ref{eqn1}), (\ref{eqn23}), and the generalized Friedmann equation (\ref{eqn26}), we can derive a first-order differential equation for the average Hubble parameter $H$ in the following form:
\begin{equation}
\overset{.}{H}+\left[ \frac{9\alpha -\left( n+2\right) ^{2}\zeta _{1}}{%
2\alpha \left( n+2\right) }\right] H^{2}-\frac{\left( n+2\right) \zeta _{0}}{%
2\alpha }H=0.  \label{eqn28}
\end{equation}%

It is important to note that, to reduce the number of model parameters, we made the assumption $\zeta_2=0$ \cite{Solanki3}. Specifically, when the viscosity coefficient depends on velocity but not on acceleration. Now, we substitute the term $\frac{d}{dt}$ with $\frac{d}{d\ln(a)}$ using the expression $\frac{d}{dt}=H \frac{d}{d\ln(a)}$, resulting in Eq. \eqref{eqn28} taking the form:
\begin{equation}
\frac{dH}{d\ln(a)}+\left[ \frac{9\alpha -\left( n+2\right) ^{2}\zeta _{1}}{%
2\alpha \left( n+2\right) }\right] H-\frac{\left( n+2\right) \zeta _{0}}{%
2\alpha }=0.  \label{eqn29}
\end{equation}%

Then, we consider a cosmological source that emits light, and as a result of cosmic expansion, the emitted light experiences a redshift. The redshift $z$ due to the expansion of the Universe is expressed as $1+z=\frac{a_{0}}{a\left( t\right) }$, where $a(t)$ represents the scale factor at the time when the object emitted the light reaching us, and $a_{0}$denotes the current value of the scale factor, which we take as $a_{0}=1$. Therefore, the integration of Eq. (\ref{eqn29}) yields the following solution for the Hubble parameter in terms of redshift:
\begin{widetext}
\begin{equation}
H(z)=\frac{(z+1)^{-\frac{\zeta_1 (n+2)}{2 \alpha }} \left[H_0 (z+1)^{\frac{9}{2 (n+2)}} \left(9 \alpha -\zeta_1 (n+2)^2\right)-\zeta_0 (n+2)^2 \left((z+1)^{\frac{9}{2 (n+2)}}-(z+1)^{\frac{\zeta_1 (n+2)}{2 \alpha }}\right)\right]}{9 \alpha -\zeta_1 (n+2)^2},  \label{eqn30}
\end{equation}
\end{widetext}
where $H(0)=H_0$ represents the current value of the Hubble parameter. In particular, when $\zeta_0=\zeta_1=0$, and $\alpha=n=1$, the expression for the Hubble parameter $H(z)$, simplifies to $H(z) = H_0 (1 + z)^{\frac{3}{2}}$. This specific configuration corresponds to the non-viscous matter-dominated Universe. In this scenario, the absence of bulk viscosity, coupled with specific choices for $\alpha$ and $n$, results in the classical evolution where the Hubble parameter follows the $(1 + z)^{\frac{3}{2}}$ scaling.

\section{Observational Limitations and Constraints}
\label{sec4}

To investigate the observational characteristics of our cosmological model, we leverage the latest cosmic Hubble, SN observations, and BAO. Our dataset includes 31 points from the Hubble dataset, 1048 points from the Pantheon SN samples, and 6 points from the BAO dataset. Employing Bayesian analysis, we utilize the likelihood function and the Markov Chain Monte Carlo (MCMC) method implemented in the \texttt{emcee} Python library \cite{Mackey/2013}.

\subsection{Hubble dataset}

In the field of observational cosmology, the universe's expansion can be directly studied through the Hubble parameter, denoted as $H =\frac{\Dot{a}}{a}$, where $\Dot{a}$ represents the derivative of the cosmic scale factor $a(t)$ with respect to cosmic time $t$. The Hubble parameter as a function of redshift can be represented as $H(z)=-\frac{dz}{dt(1+z)}$. Given that $dz$ is obtained from a spectroscopic survey, the model-independent value of the Hubble parameter can be computed by measuring the quantity $dt$. We include a set of 31 data points, measured through the differential age approach \cite{Sharov/2018}, to avoid additional correlations with BAO data. The mean values of the model parameters $H_0$, $\alpha$, $n$, $\zeta_0$, and $\zeta_1$ are calculated using the chi-square function as follows:
\begin{equation}
\chi _{H}^{2}(H_0,\alpha,n,\zeta_0,\zeta_1)=\sum\limits_{k=1}^{31}
\frac{[H_{th}(H_0,\alpha,n,\zeta_0,\zeta_1, z_{k})-H_{obs}(z_{k})]^{2}}{
\sigma _{H(z_{k})}^{2}}, 
\end{equation}
where $H_{th}$ denotes the predicted Hubble parameter value from the model, $H_{obs}$ represents its observed value, and the standard error in the observed value of $H$ is denoted by $\sigma_ {H(z{k})}$.

\subsection{Pantheon dataset}

At first, studies of Type Ia SNe, using a key sample of 50 points, suggested that the universe is expanding at an accelerating rate. Over the last two decades, research in this area has expanded, incorporating larger and larger datasets of supernovae. A recent milestone is the release of a new dataset containing 1048 data points from Type Ia SNe. Recently, Scolnic et al. \cite{Scolnic/2018} compiled the Pantheon samples, encompassing 1048 SNe within the redshift range $0.01 < z < 2.3$. The PanSTARSS1 Medium Deep Survey, SDSS, SNLS, and various low-z and HST samples contribute to this dataset. The empirical relation employed for calculating the distance modulus of SNeIa from the observation of light curves is expressed as $\mu= m_{B}^{*}+\alpha X_{1}-\beta C-M_{B} + \Delta_{M}+ \Delta_{B}$, where $X_{1}$ and $C$ denote the stretch and color correction parameters, respectively \cite{Scolnic/2018}. Here, $m_{B}^*$ represents the observed apparent magnitude, and $M_{B}$ is the absolute magnitude in the B-band for SNe. The parameters $\alpha$ and $\beta$ serve as two nuisance parameters characterizing the luminosity stretch and luminosity color relations, respectively. Furthermore, there is a distance correction factor denoted by $\Delta_{M}$, and $\Delta_{B}$ represents a distance correction based on anticipated biases from simulations. The nuisance parameters in the Tripp formula \cite{Tripp/1998} were determined using a novel technique called BEAMS with Bias Corrections \cite{Kessler/2017, Fotios/2021}. The observed distance modulus was then reduced to the difference between the corrected apparent magnitude $m_{B}$ and the absolute magnitude $M_{B}$, defined as $\mu= m_{B}-M_{B}$. In our current investigation of the model, we choose to avoid marginalizing over the nuisance parameters $\alpha$ and $\beta$ but instead marginalize over the Pantheon data for $M_{B}$. Therefore, we exclude the values of $\alpha$ and $\beta$ from the present analysis.

The luminosity distance is expressed as
\begin{equation}
D_{L}(z)= c(1+z) \int_{0}^{z} \frac{dz'}{H(z')},
\end{equation}
where $c$ represents the speed of light. Furthermore, the theoretical distance modulus is
\begin{equation}
\mu(z)= 5log_{10}D_{L}(z)+\mu_{0}, 
\end{equation}
where
\begin{equation}
\mu_{0} =  5log(1/H_{0}Mpc) + 25.
\end{equation}

In the case of the Pantheon dataset, the mean values for the model parameters $H_0$, $\alpha$, $n$, $\zeta_0$, and $\zeta_1$ are determined through the chi-square function as follows:
\begin{equation}\label{4i}
\chi^{2}_{SN}(H_0,\alpha,n,\zeta_0,\zeta_1)= \sum _{k=1}^{1048} \dfrac{\left[ \mu_{obs}(z_{k})-\mu_{th}(H_0,\alpha,n,\zeta_0,\zeta_1, z_{k})\right] ^{2}}{\sigma^{2}(z_{k})},
\end{equation}
where $\mu_{th}$ represents the theoretical value of the distance modulus, $\mu_{obs}$ denotes the observed value, and $\sigma^{2}(z_{k})$ signifies the standard error in the observed value.

\subsection{BAO dataset}

When investigating the early universe, baryons, photons, and DM form a unified fluid that is tightly coupled through Thomson scattering. Despite the presence of gravity, this fluid does not collapse; instead, it oscillates due to the significant pressure exerted by photons. BAO is an analytical framework that specifically addresses these oscillations during the early stages of the Universe. Here, we use the BAO distance dataset, which consists of measurements from the 6dFGS, SDSS, and WiggleZ surveys, providing BAO measurements at six distinct redshifts. The characteristic scale of BAO is determined by the sound horizon $r_{s}$ at the epoch of photon decoupling $z_{\ast }$, as given by the following relation:
\begin{equation}\label{4b}
r_{s}(z_{\ast })=\frac{c}{\sqrt{3}}\int_{0}^{\frac{1}{1+z_{\ast }}}\frac{da}{
a^{2}H(a)\sqrt{1+(3\Omega _{b0}/4\Omega _{\gamma 0})a}}.
\end{equation}

In this context, $\Omega _{b0}$ and $\Omega _{\gamma 0}$ represent the present densities of baryons and photons, respectively. In addition, the BAO measurements utilize the following relations:
\begin{equation}
 \triangle \theta =\frac{r_{s}}{d_{A}(z)},
\end{equation}
\begin{eqnarray}
d_{A}(z)=\int_{0}^{z}\frac{dz^{\prime }}{H(z^{\prime })}, 
\end{eqnarray}  
\begin{equation}
	\triangle z=H(z)r_{s},
\end{equation}
where $\triangle \theta $ signifies the measured angular separation, $d_{A}$ represents the angular diameter distance, and $\triangle z$ denotes the measured redshift separation of the BAO feature in the 2-point correlation function of the galaxy distribution on the sky along the line of sight. In this study, we utilize a BAO dataset consisting of six points for $d_{A}(z_{\ast })/D_{V}(z_{BAO})$, obtained from the references \cite{BAO1, BAO2, BAO3, BAO4, BAO5, BAO6}. Here, the redshift at the epoch of photon decoupling is considered as $z_{\ast }\approx 1091$, and $d_{A}(z)$ represents the co-moving angular diameter distance along with the dilation scale $D_{V}(z)=\left[d_{A}(z)^{2}z/H(z)\right] ^{1/3}$. The chi-square function for the BAO dataset is given by \cite{BAO6},
\begin{equation}
\chi _{BAO}^{2}=X^{T}C^{-1}X\,,  
\end{equation}
where the variable $X$ is contingent on the specific survey under consideration, and $C$ represents the covariance matrix \cite{BAO6}.

From the Hubble, Pantheon, and BAO datasets, we extract the best-fit values for $H_0$, $\alpha$, $n$, $\zeta_0$, and $\zeta_1$ as illustrated by the $1-\sigma$ and $2-\sigma$ contour plots in Fig. \ref{Combine}. The determined best-fit values are presented in Tab. \ref{tab}. Fig. \ref{Hubblecond} displays the error bar plot for the considered model in comparison to the $\Lambda$CDM or standard cosmological model, where the cosmological constant density parameter is denoted as $\Omega_{\Lambda} = 0.685$, the matter density parameter as $\Omega_{m_0} = 0.315$, and $H_0 = 67.4$ $km/s/Mpc$ \cite{ref49}.

\subsection{Information-based criteria for model selection analysis}

To evaluate the effectiveness of our MCMC study, we need to conduct a statistical assessment using the Akaike Information Criterion (AIC) and Bayesian Information Criterion (BIC). The AIC can be formulated as \cite{Akaike}
\begin{equation}
    \mathrm{AIC} = \chi^2_{\mathrm{min}}+2d,
\end{equation}
where $d$ represents the count of independent parameters in the model under consideration. Moreover, BIC is determined by
\begin{equation}
\mathrm{BIC} =\chi^2_{\mathrm{min}}+d\ln N.
\end{equation}

Here, $N$ represents the number of data points utilized in MCMC. For comparing our outcomes with the standard $\Lambda$CDM model, we utilize the AIC difference between our model and the standard cosmological model,
\begin{equation}
    \Delta\mathrm{AIC}=\mathrm{AIC}_{\Lambda\mathrm{CDM}}-\mathrm{AIC}_{\mathrm{Model}}.
\end{equation}

In this context, when the difference ($\Delta\mathrm{AIC}$ and/or $\Delta\mathrm{BIC}$) is below 2, it suggests "consistency" between the compared models. A difference in the range of 2 to 6 indicates "positive evidence" in favor of the model with the smaller $\Delta\mathrm{AIC}$ and/or $\Delta\mathrm{BIC}$ value. Differences falling within 6 to 10 suggest "strong evidence" in favor of the chosen model. If the difference exceeds 10, it is considered "very strong evidence" in favor of the model \cite{Liddle}. Consequently, we have compiled the $\chi^2_{\mathrm{min}}$, AIC, and $\Delta\mathrm{AIC}$ data for our model in Tab. \ref{tab2}. Upon reviewing the values, we observe that the data provides moderate support for our model. Specifically, a result of 1.5 ($Hubble$) indicates consistency rather than support, as does a result of 1.8 ($SN$), while a result of 2.5 ($BAO$) would only mildly support the model.

\begin{widetext}

\begin{figure}[H]
\centerline{\includegraphics[scale=0.3]{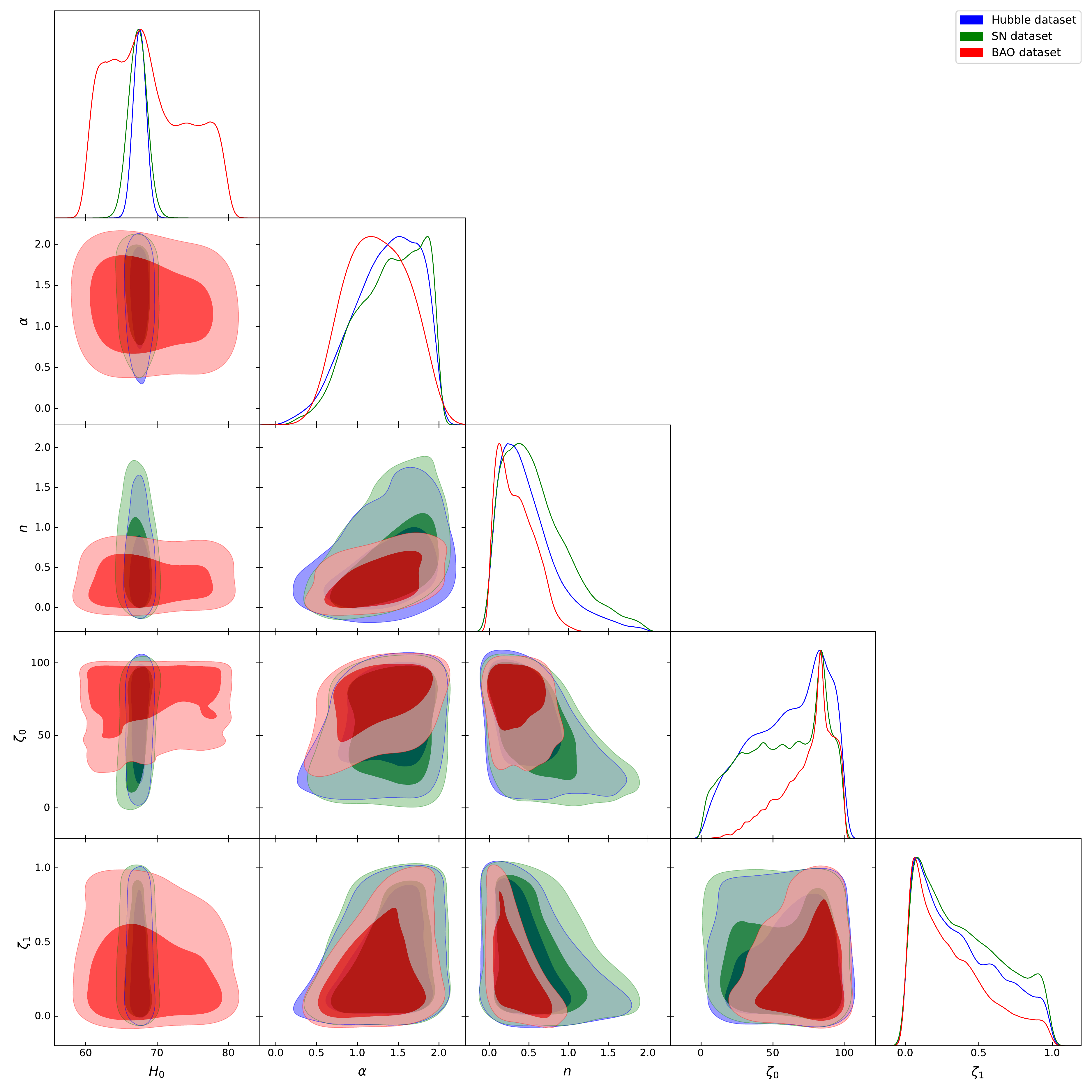}}
\caption{Contour plot: Joint likelihood function for model parameters $H_0$, $\alpha$, $n$, $\zeta_0$, and $\zeta_1$ using $Hubble$, $SN$, and $BAO$ data with $1-\sigma $ and $2-\sigma $ confidence levels.}
\label{Combine}
\end{figure}

\begin{figure}[H]
\centerline{\includegraphics[scale=0.5]{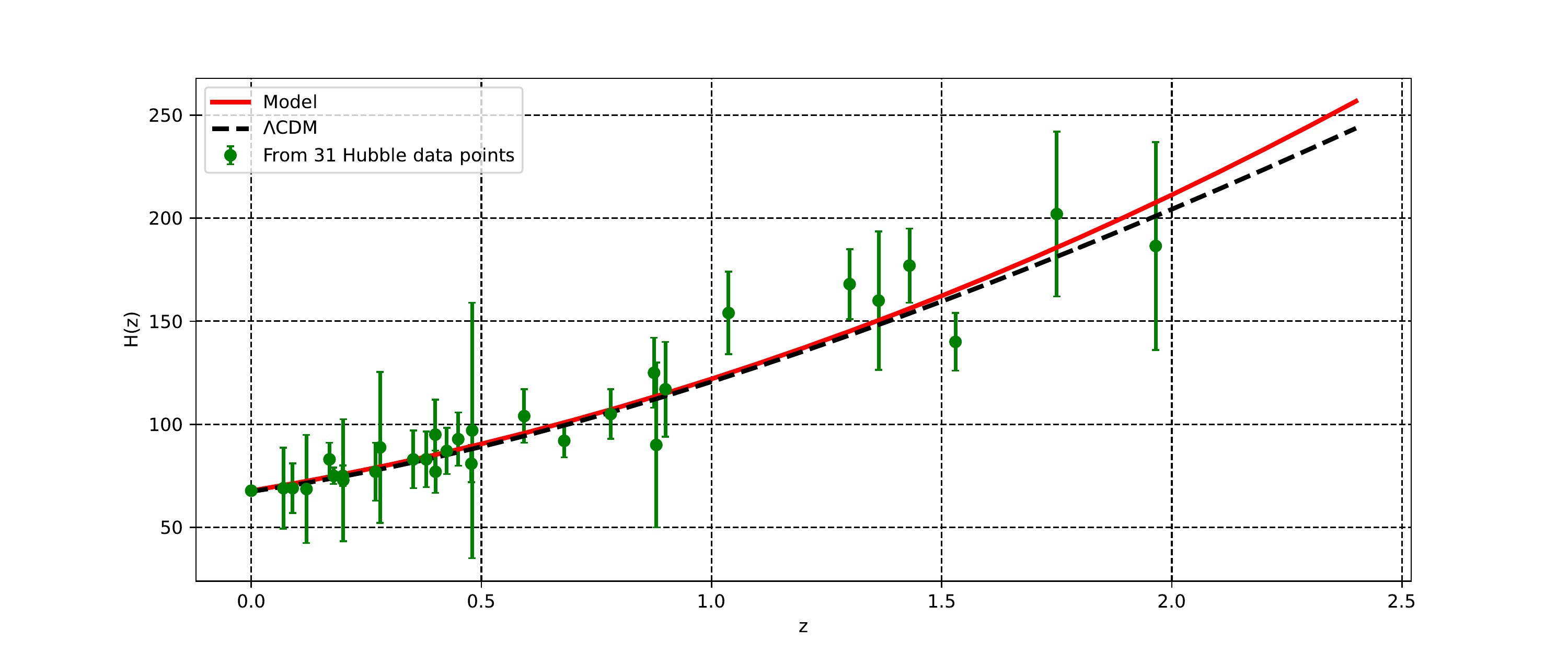}}
\caption{Comparison between the model and $\Lambda$CDM for the Hubble parameter $H(z)$ as a function of redshift $z$. The Model curve is depicted by the green line, while the $\Lambda$CDM model is represented by the black dotted line. The green dots with error bars correspond to the 31 Hubble sample points.}
\label{Hubblecond}
\end{figure}

\begin{table*}[h]
\begin{center}
\begin{tabular}{c c c c c c}
\hline\hline 
$Datasets$              & $H_{0}$ $(km/s/Mpc)$ & $\alpha$ & $n$ & $\zeta_{0}$ & $\zeta_{1}$\\
\hline
$Priors$   & $(60,80)$  & $(0,2)$  & $(0,2)$ & $(0,100)$ & $(0,1)$\\

$Hubble$ & $67.5^{+1.7}_{-1.7}$ & $1.35^{+0.66}_{-0.77}$  & $0.50^{+0.81}_{-0.55}$ & $61^{+40}_{-50}$ & $0.37^{+0.53}_{-0.38}$\\

$SN$   & $67.3^{+2.5}_{-2.5}$ & $1.40^{+0.63}_{-0.73}$  & $0.60^{+0.88}_{-0.63}$ & $57^{+40}_{-50}$ & $0.40^{+0.53}_{-0.40}$\\

$BAO$   & $69^{+10}_{-8}$ & $1.25^{+0.69}_{-0.69}$  & $0.34^{+0.44}_{-0.35}$ & $74^{+30}_{-40}$ & $0.32^{+0.54}_{-0.34}$\\

\hline\hline
\end{tabular}
\caption{The marginalized constraints for the parameters $H_0$, $\alpha$, $n$, $\zeta_{0}$, and $\zeta_{1}$ are presented for various data samples at 68\% and 95\% confidence levels.}
\label{tab}
\end{center}
\end{table*}

\begin{table*}[h]
\begin{center}
\begin{tabular}{c c c c }
\hline\hline
				$Model$ & $\chi^2_{min}$ & $AIC$ & $\Delta AIC$ \\
    	\hline
     & $Hubble$ & &\\
    $\Lambda CDM$ & 22.028 & 26.028 & 0\\
    $f(Q)$ gravity & 14.739 & 24.739 & 1.3\\

     & $SN$ & &\\
    $\Lambda CDM$ & 1049.785& 1053.785 & 0\\
    $f(Q)$ gravity & 1041.996 & 1051.996 & 1.8\\

     & $BAO$ & &\\
   $\Lambda CDM$  & 10.711 & 14.711 & 0\\
    $f(Q)$ gravity & 2.190 & 12.190 & 2.5\\
    \hline\hline

\end{tabular}
\caption{The $\chi^2_{min}$ values for each model are provided for every sample, as well as the AIC for the investigated cosmological models. Also, the differences $\Delta AIC= AIC_{\Lambda CDM}-AIC_{Model}$ are included.}
\label{tab2}
\end{center}
\end{table*}

\end{widetext}

\section{Cosmological Parameters}
\label{sec5}

\subsection{Deceleration parameter (DP)}

According to recent observations in cosmology, our Universe is presently undergoing a transition from an early decelerating phase to the current accelerating stage. Understanding the expansion of the Universe involves investigating the behavior of the Deceleration Parameter (DP), defined as
\begin{equation}
q=-1-\frac{\overset{.}{H}}{H^{2}}.  \label{eqn39}
\end{equation}

\begin{figure}[h]
\centerline{\includegraphics[scale=0.7]{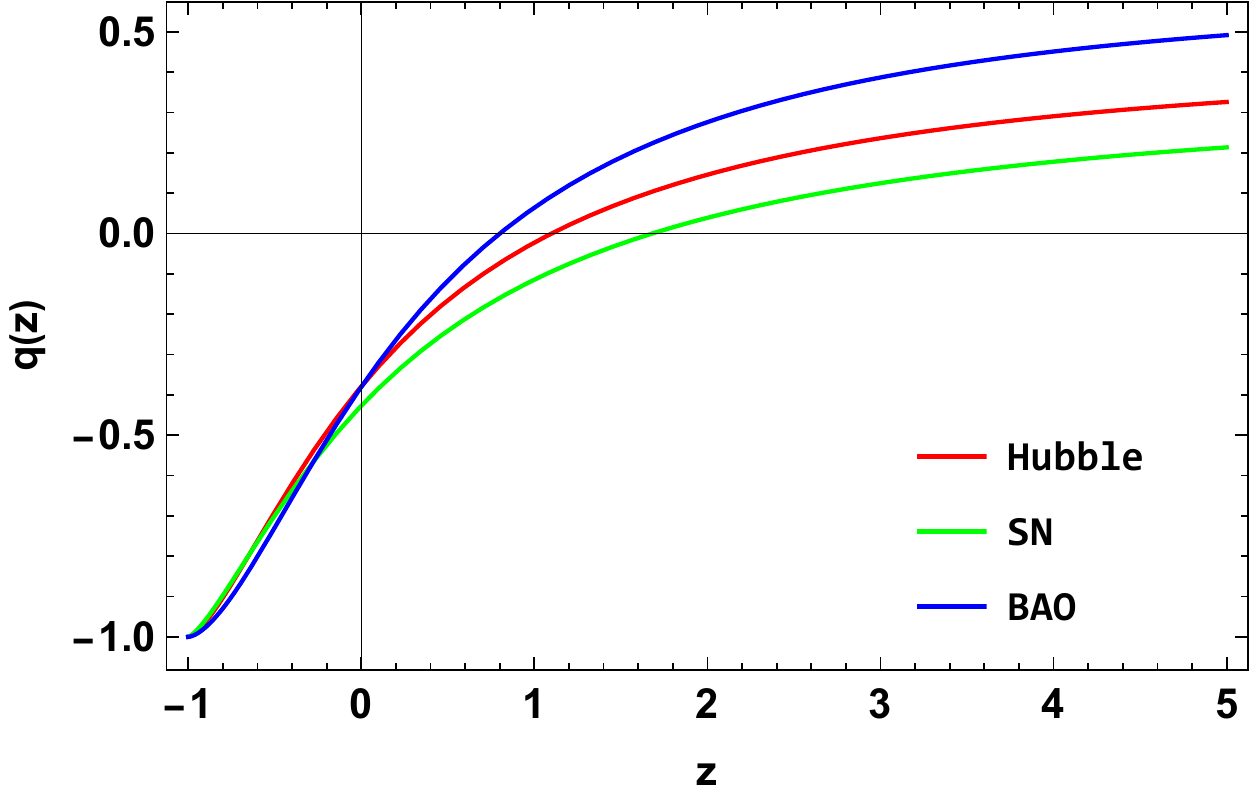}}
\caption{Deceleration parameter ($q$) versus redshift ($z$) plot for constrained parameter values from Hubble, Pantheon (SN), and BAO datasets.}
\label{fig1}
\end{figure}

This parameter is positive ($q>0$) when the Universe's expansion decelerates over time, signifying a phase of slowing cosmic expansion. Conversely, it is negative ($q<0$) when the Universe undergoes acceleration, indicating an epoch where the rate of expansion increases. In the limiting case where all distances in the Universe evolve linearly with time, the parameter takes on a value of zero, representing a critical state where the expansion neither decelerates nor accelerates. Fig. \ref{fig1} illustrates the behavior of the DP as a function of redshift, $z$, considering the Hubble, Pantheon, and BAO datasets. The plot demonstrates that our model aligns well with the expected evolution of the Universe, showcasing an early decelerating phase followed by late-time cosmic acceleration. The current value of the DP denoted as $q_{0}$, is determined to be $q_{0}=-0.38$, $q_{0}=-0.43$, and $q_{0}=-0.37$ for the Hubble, Pantheon, and BAO datasets, respectively. When considering the numerical values of $H_0$ and $\Omega_{m0}$ obtained from the latest Planck data for the standard $\Lambda$CDM model \cite{ref49}, the current value of the DP is $q_0=-0.53$, and the value predicted by our model is close to this reference value.

\subsection{EoS parameter}
In cosmology, the EoS parameter is defined as the ratio between the pressure and the energy density, expressed as $\omega =\frac{p}{\rho }$. When dealing with a Universe filled with a viscous fluid, the EoS parameter becomes crucial for characterizing its behavior. Notably, in the context of cosmic acceleration in modified theories of gravity, the EoS parameter is typically negative. This characteristic is evident when examining the Friedmann equations within the standard model; specifically, when $\rho +3p<0$, it leads to $\omega <-\frac{1}{3}$. For instance, in the case of the cosmological constant ($\Lambda$CDM), the EoS parameter is $\omega =-1$. In other components of the Universe, such as radiation, $\omega =\frac{1}{3}$, and for non-relativistic matter, $\omega =0$. The expressions for the matter-energy density and the bulk viscous pressure in our model are provided as follows:
\begin{equation}
\rho =\frac{9\alpha \left( 1+2n\right) }{\left( n+2\right) ^{2}}H^{2},
\label{eqn42}
\end{equation}
\begin{equation}
p_{v}=-\frac{27\alpha }{\left( n+2\right) ^{2}}H^{2}-\frac{6\alpha }{\left(
n+2\right) }\overset{.}{H}.
\end{equation}%

Thus, we can determine the effective EoS parameter as
\begin{equation}
\omega =-\frac{3}{\left( 1+2n\right) }-\frac{2\left( n+2\right) }{3\left(
1+2n\right) }\frac{\overset{.}{H}}{H^{2}}.  \label{eqn44}
\end{equation}

\begin{figure}[h]
  \centerline{\includegraphics[scale=0.7]{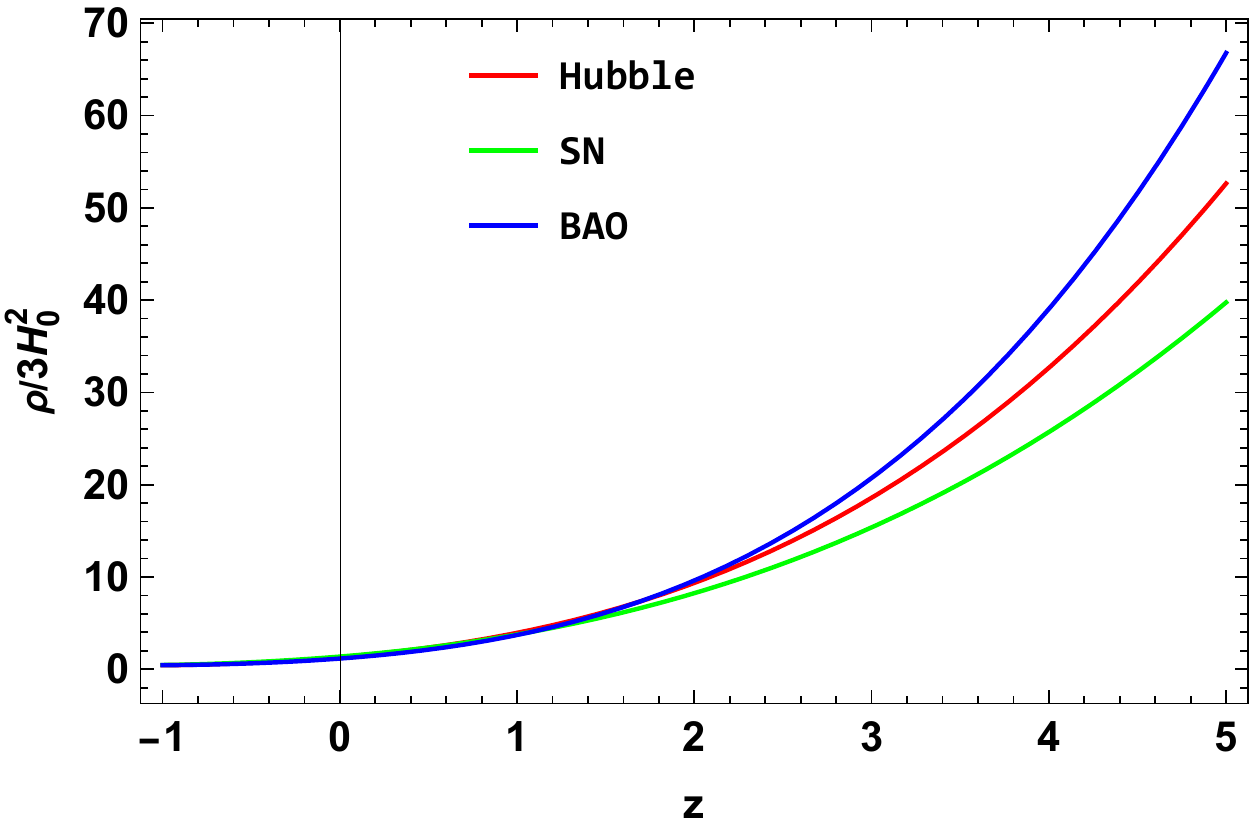}}
  \caption{Matter-energy density ($\rho$) versus redshift ($z$) plot for constrained parameter values from Hubble, Pantheon (SN), and BAO datasets.}\label{fig2}
 \end{figure}
 
\begin{figure}[h]
  \centerline{\includegraphics[scale=0.7]{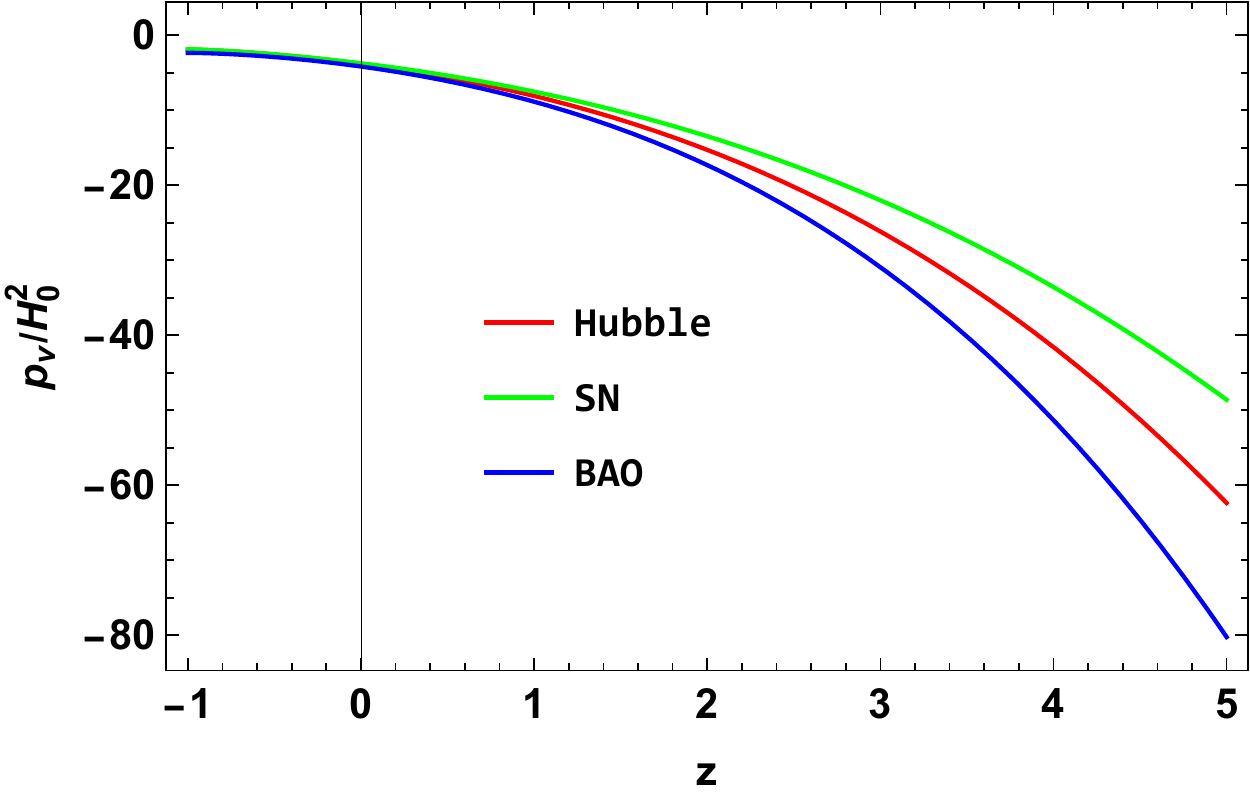}}
  \caption{Bulk viscous pressure ($p_v$) versus redshift ($z$) plot for constrained parameter values from Hubble, Pantheon (SN), and BAO datasets.}\label{fig3}
\end{figure}

\begin{figure}[h]
\centerline{\includegraphics[scale=0.7]{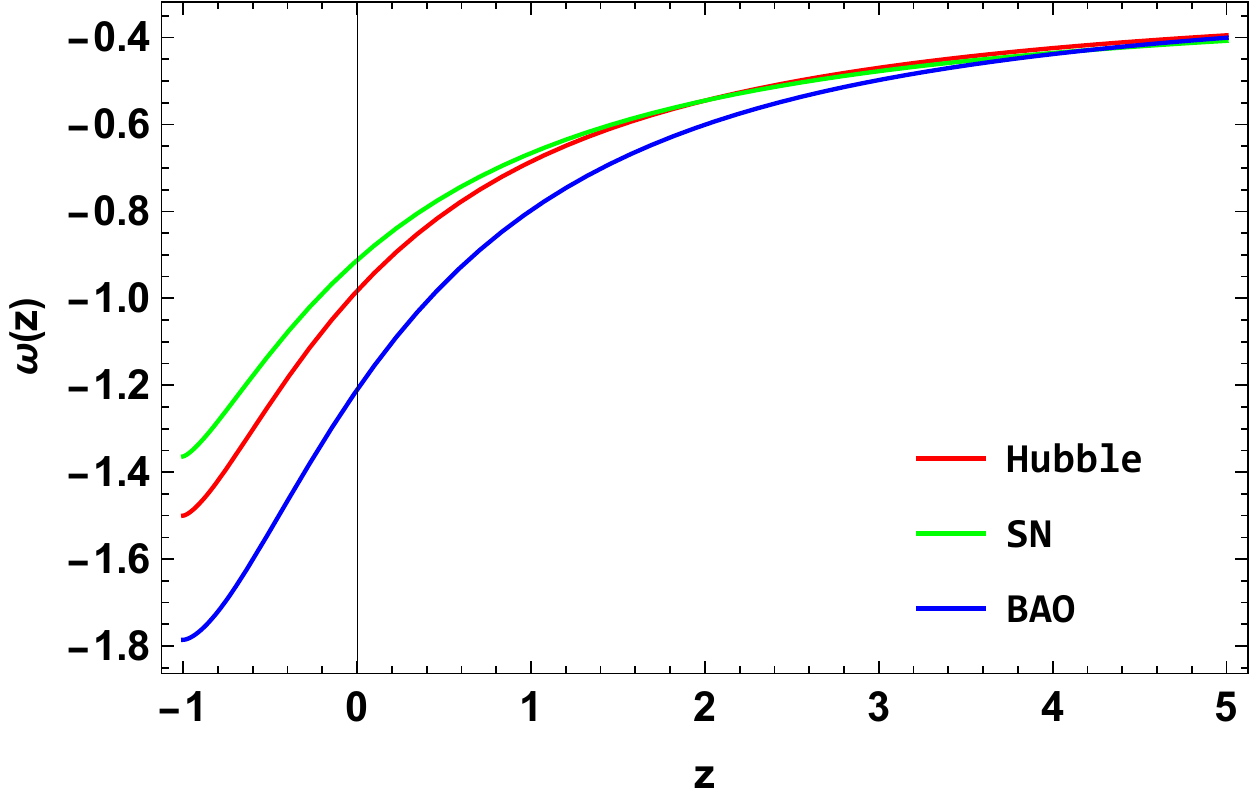}}
\caption{Effective EoS parameter ($\omega$) versus redshift ($z$) plot for constrained parameter values from Hubble, Pantheon (SN), and BAO datasets.}
\label{fig4}
\end{figure}

Fig. \ref{fig2} clearly illustrates that the matter-energy density is an increasing function of the redshift parameter $z$ and approaches a small value in the distant future (i.e., as $z\rightarrow -1$) for all the constrained values of the model parameters. In contrast, the bulk viscous pressure, as shown in Fig. \ref{fig3}, is a decreasing function of redshift $z$ and maintains negative values throughout the cosmic evolution. It starts with enormous negative values in the Universe's early stages and gradually approaches zero over time. The presence of negative bulk viscous pressure aligns with the accelerating phase of the Universe, which is consistent with recent observational evidence, confirming the validity of our model. In Fig. \ref{fig4}, the behavior of the effective EoS parameter is depicted as a function of redshift $z$ for the Hubble, Pantheon, and BAO datasets. Notably, the behavior of the effective EoS parameter resembles that of the quintessence DE model, falling within the range $-1<\omega <-\frac{1}{3}$ for Hubble and Pantheon datasets, while for the BAO dataset, it exhibits phantom behavior with $\omega <-1$. Furthermore, the present value of the effective EoS parameter is determined as $\omega _{0}=-0.98$, $\omega _{0}=-0.92$, $\omega _{0}=-1.21$ for the Hubble, Pantheon, and BAO datasets, respectively. These present values of $\omega $ are consistent with the results found in the literature \cite{Mukherjee1,Novosyadlyj/2012,Suresh/2014}.

\subsection{Skewness parameter}
By definition, the skewness parameter quantifies the degree of anisotropy present in a DE bulk viscous fluid. It is denoted as $\delta $, and by utilizing Eqs. (\ref{eqn25}), (\ref{eqn26}), and (\ref{eqn27}), we can express it as follows:
\begin{equation}
\delta =-\frac{\left( n^{2}+n-2\right) }{\left( 1+2n\right) }-\frac{\left(
n+2\right) \left( n-1\right) }{3\left( 1+2n\right) }\frac{\overset{.}{H}}{%
H^{2}}.  \label{eqn45}
\end{equation}

\begin{figure}[h]
\centerline{\includegraphics[scale=0.7]{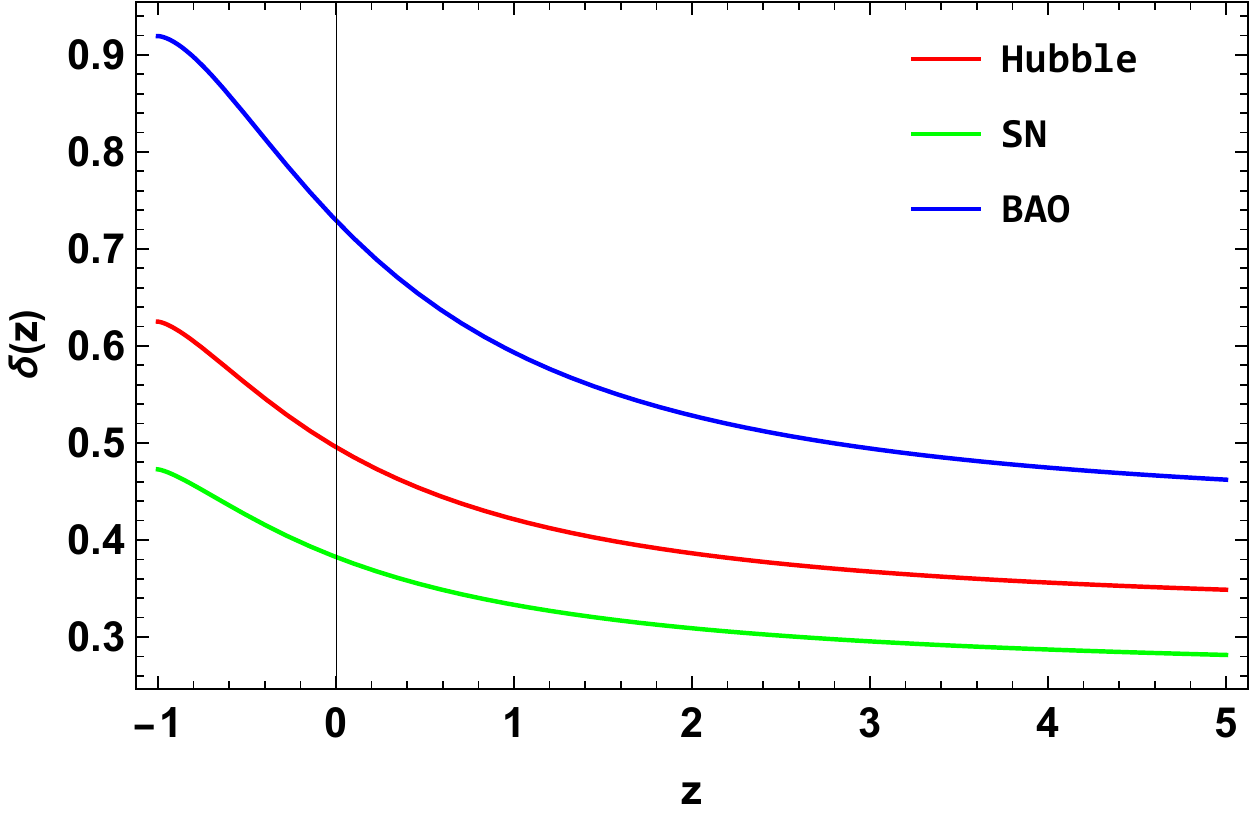}}
\caption{Skewness parameter ($\delta$) versus redshift ($z$) plot for constrained parameter values from Hubble, Pantheon (SN), and BAO datasets.}
\label{fig5}
\end{figure}

Fig. \ref{fig5} illustrates the behavior of the skewness parameter as a function of the redshift $z$ for the Hubble, Pantheon, and BAO datasets. 
It is evident from this figure that the skewness parameter assumes positive values both in the past and the future, as well as at the present epoch. Therefore, we can deduce that our model exhibits anisotropy throughout the evolution of the Universe.

\subsection{Statefinder diagnostic}

As researchers have increasingly focused on addressing the issue of cosmic acceleration and DE, numerous models of DE have emerged, making it challenging to navigate and differentiate among them. Sahni et al. \cite{ref50} introduced a novel concept known as the statefinder diagnostic, denoted as $(r,s)$, which constitutes a geometrical parameter primarily designed to elucidate cosmic acceleration while distinguishing between various DE models. The statefinder parameters $(r,s)$ are expressed in terms of the Hubble parameter and the DP as follows:
\begin{equation}
r=\frac{\overset{...}{a}}{aH^{3}},\text{ \ \ }s=\frac{r-1}{3\left( q-\frac{1%
}{2}\right) }.  \label{eqn46}
\end{equation}

\begin{figure}[h]
\centerline{\includegraphics[scale=0.7]{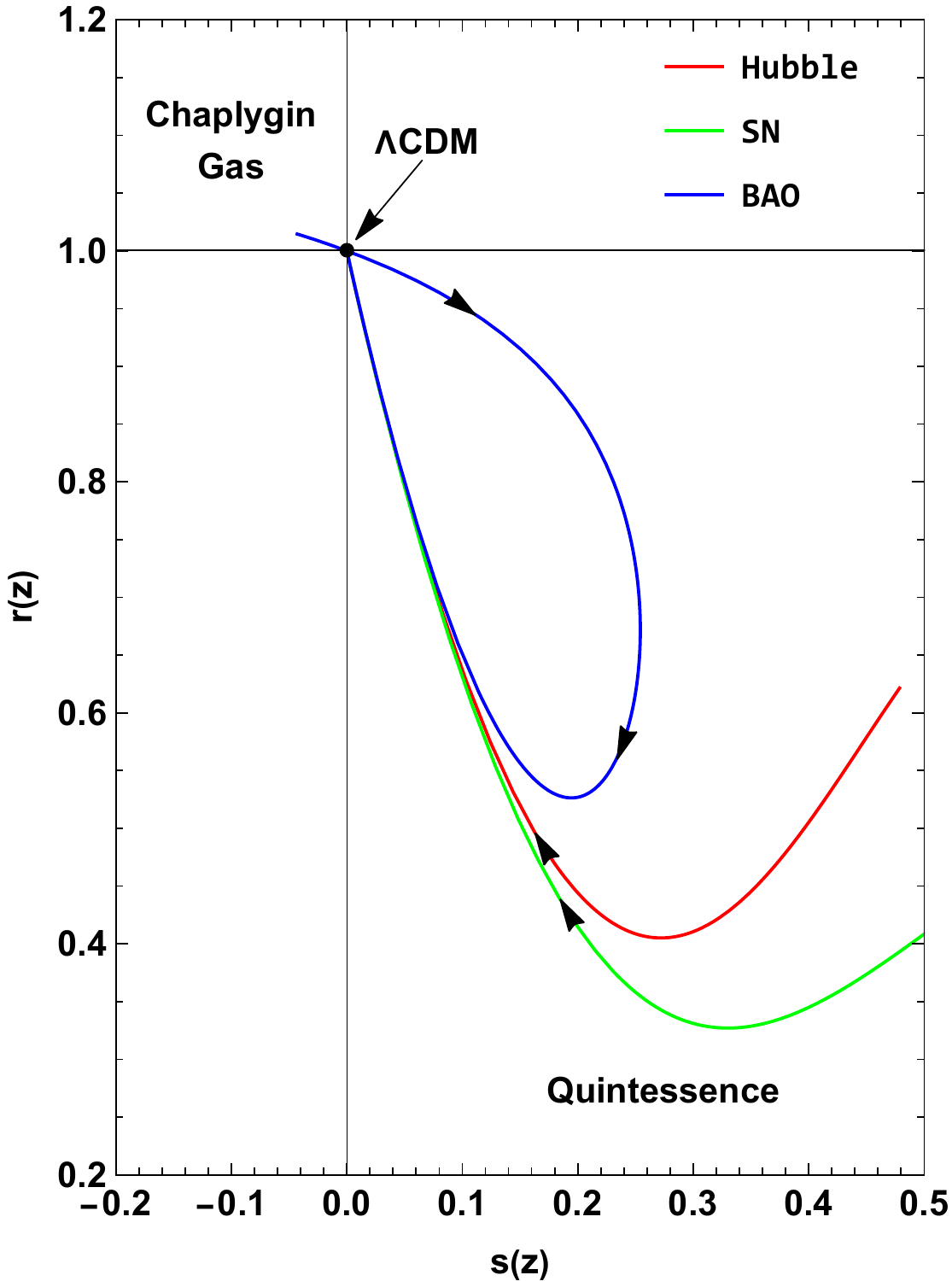}}
\caption{Statefinder parameters $(r,s)$ plot for constrained parameter values from Hubble, Pantheon (SN), and BAO datasets.}
\label{fig6}
\end{figure}

The statefinder parameters $(r,s)$ are associated with distinct regions of cosmological interest. When $(r,s) =(1,1)$, it corresponds to the standard cold dark matter (SCDM) limit, while the fixed point $(r,s)=(1,0)$ corresponds to the spatially flat $\Lambda$CDM limit. For values where $r<1$ and $s>0$, it characterizes regions related to DE, including quintessence and phantom eras. In this paper, we illustrate the evolution of the statefinder parameters $(r, s)$ in Fig. \ref{fig6} for constrained parameter values from the Hubble, Pantheon, and BAO datasets. It is noteworthy that the statefinder parameters $(r, s)$ for a Universe governed by a bulk viscous fluid initially exhibit behavior reminiscent of the quintessence model ($r<1$ and $s>0$), with a slight deviation from the quintessence model for the BAO dataset. However, as time progresses, they tend to approach the characteristics of the $\Lambda$CDM model ($r=1$ and $s=0$) in the future.

\subsection{$Om(z)$ diagnostic}

\begin{figure}[h]
\centerline{\includegraphics[scale=0.7]{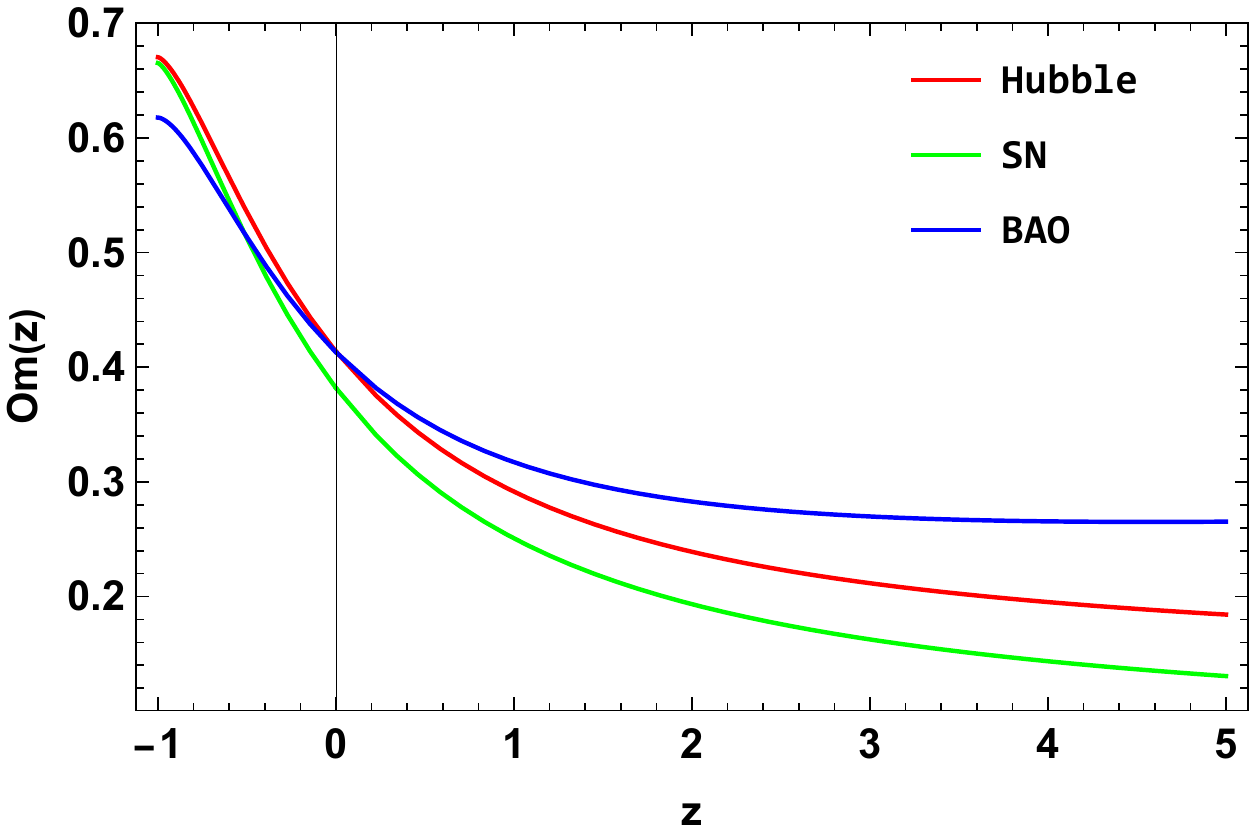}}
\caption{$Om(z)$ diagnostic versus redshift ($z$) plot for constrained parameter values from Hubble, Pantheon (SN), and BAO data sets.}
\label{fig7}
\end{figure}

Besides the statefinder parameters $(r, s)$, another diagnostic tool commonly employed in the literature for distinguishing between DE models and gaining deeper insights into constructed cosmological models is the $Om(z)$ diagnostic. This diagnostic is a function of the Hubble parameter and the redshift $z$ \cite{ref51}. The $Om(z)$ diagnostic for a flat Universe is calculated as:
\begin{equation}
Om\left( z\right) =\frac{\left( \frac{H\left( z\right) }{H_{0}}\right) ^{2}-1%
}{\left( 1+z\right) ^{3}-1},  \label{eqn49}
\end{equation}%
where $H_0$ represents the present value of the Hubble parameter. Similar to the statefinder parameters $(r, s)$, the $Om(z)$ diagnostic assumes different values depending on the characteristics of the investigated Dark Energy (DE) model. Consequently, positive, negative, and constant slopes of $Om(z)$ correspond to the phantom ($\omega_{DE} <-1$), quintessence ($\omega_{DE} >-1$), and flat $\Lambda$CDM ($\omega_{DE}=-1$) DE models, respectively. In Fig. \ref{fig7}, it is evident that the $Om(z)$ diagnostic assumes a negative slope, indicating the similarity of our model to the quintessence scenario.

\section{Conclusions}
\label{sec6}

In the context of hydrodynamics, the consideration of viscous effects in the cosmic fluid is inherently natural, as the idealization of a perfect fluid is, in reality, an abstraction. In astrophysical and cosmological contexts, real fluids often exhibit non-ideal behavior, and the inclusion of bulk viscosity is a plausible extension to capture more realistic aspects of the cosmic medium. Bulk viscosity accounts for dissipative effects on large scales, reflecting the interactions and complexities present in the cosmic fluid, especially during phases of cosmic evolution where deviations from idealized behavior become significant \cite{ref43,ref44,ref45}. In this study, we have explored the dynamics of the LRS B-I Universe in the presence of non-relativistic bulk viscous matter within the framework of $f(Q)$ symmetric teleparallel gravity. Utilizing a specific form for the bulk viscous coefficient, $\zeta = \zeta_0 + \zeta_1 H$, where $H$ represents the Hubble parameter, and $\zeta_0$ and $\zeta_1$ are constants \cite{ref44}, we derived exact solutions for the field equations governing the Universe under the influence of bulk viscosity. We considered a linear $f(Q) = \alpha Q$ model, where $\alpha \neq 0$ serves as a free parameter, and introduced an additional constraint $H_x = nH_y$ (where $n\neq 0,1$ is an arbitrary real number) to facilitate the determination of the Hubble parameter.

Moreover, we assessed the validity of the proposed 
$f(Q)$ model by incorporating observational datasets, specifically the Hubble dataset, Pantheon SNe dataset, and BAO dataset. The resulting best-fit values are as follows: for the Hubble dataset, $H_0=67.5^{+1.7}_{-1.7}$ $km/s/Mpc$, $\alpha=1.35^{+0.66}_{-0.77}$, $n=0.50^{+0.81}_{-0.55}$, $\zeta_0=61^{+40}_{-50}$, and $\zeta_1=0.37^{+0.53}_{-0.38}$; for the Pantheon dataset, $H_0=67.3^{+2.5}_{-2.5}$ $km/s/Mpc$, $\alpha=1.40^{+0.63}_{-0.73}$, $n=0.60^{+0.88}_{-0.63}$, $\zeta_0=57^{+40}_{-50}$, and $\zeta_1=0.40^{+0.53}_{-0.40}$; and for the BAO dataset, $H_0=69^{+10}_{-8}$ $km/s/Mpc$, $\alpha=1.25^{+0.69}_{-0.69}$, $n=0.34^{+0.44}_{-0.35}$, $\zeta_0=74^{+30}_{-40}$, and $\zeta_1=0.32^{+0.54}_{-0.34}$. The main findings can be summarized as follows: Our model demonstrates a decelerating phase in the early Universe, transitioning to an accelerating phase in the present epoch, consistent with recent measurements (see Fig. \ref{fig1}). The matter-energy density remains positive but decreases over time, while the viscous pressure takes on negative values, indicative of cosmic acceleration (see Figs. \ref{fig2} and \ref{fig3}). The behavior of the effective EoS parameter closely resembles that of the quintessence model for the Hubble and Pantheon datasets. However, for the BAO dataset, it exhibits phantom behavior, and its current values (i.e. $\omega _{0}=-0.98$, $\omega _{0}=-0.92$, $\omega _{0}=-1.21$ for the Hubble, Pantheon, and BAO datasets, respectively) are in agreement with observational data (see Fig. \ref{fig4}). Further, the skewness parameter depicted in Fig. \ref{fig5} supports an anisotropic evolution of the Universe throughout its entire timeline.

Furthermore, both the statefinder parameter and the $Om(z)$ diagnostic suggest that our model shares similarities with the quintessence model in the present era, with indications of an eventual convergence towards the $\Lambda$CDM model in the future (see Figs. \ref{fig6} and \ref{fig7}).\newline

\textbf{Data availability} All data used in this study are cited in the references and were obtained from publicly available sources.

\end{document}